\documentclass{elsart}

 \usepackage{graphicx}
\usepackage{amssymb}
\usepackage{amsmath}

\begin{document}

\begin{frontmatter}



\title{Dynamics of allosteric action in multisite protein modification.}


\author[Trieste,INFN]{Edoardo Milotti\corauthref{cor}}
\ead{milotti@ts.infn.it}
\corauth[cor]{Corresponding author}
\author[Trieste,INFN,Verona2]{Alessio Del Fabbro}
\ead{delfabbro@ts.infn.it}
\author[Verona]{Chiara Dalla Pellegrina}
\ead{dallapel@sci.univr.it}
\author[Verona,INFN]{Roberto Chignola}
\ead{roberto.chignola@univr.it}
\address[Trieste]{Dipartimento di Fisica, Universit\`a di Trieste \\ Via Valerio, 2 -- I-34127 Trieste, Italy }
\address[INFN]{I.N.F.N. -- Sezione di Trieste}
\address[Verona]{Dipartimento Scientifico e Tecnologico, Facolt\`a di Scienze MM.FF.NN. \\ Universit\`a di Verona, Strada Le Grazie, 15 - CV1, I-37134 Verona, Italy}
\address[Verona2]{Facolt\`a di Scienze Motorie,  \\ Universit\`a di Verona, I-37134 Verona, Italy}

\begin{abstract}
Protein functions in cells may be activated or modified by the attachment of several kinds of chemical groups. While protein phosphorylation, i.e. the attachment of a phosphoryl (PO$_3^-$) group, is the most studied form of protein modification, and is known to regulate the functions of many proteins, protein behavior can also be modified by nitrosylation, acetylation, methylation, etc.
A protein can have multiple modification sites, and display some form of transition only when enough sites are modified. In a previous paper we have modeled the generic equilibrium properties of multisite protein modification (R.Chignola, C. Dalla Pellegrina, A. Del Fabbro, E.Milotti, Physica A {\bf 371}, 463 (2006) ) and we have shown that it can account both for sharp, robust thresholds and for information transfer between processes with widely separated timescales.
Here we use the same concepts to expand that analysis starting from a dynamical description of  multisite modification: we give analytical results for the basic dynamics and numerical results in an example where the modification chain is cascaded with a Michaelis-Menten step.
We modify the dynamics and analyze an example with realistic phosphorylation/dephosphorylation steps, and give numerical evidence of the independence of the allosteric effect from the details of the attachment-detachment processes. 
We conclude that multisite protein modification is dynamically equivalent to the classic allosteric effect.
\end{abstract}

\end{frontmatter}

%PACS
%82.39.Fk Enzyme kinetics
%87.16.Yc Regulatory chemical networks
%87.17.-d Cellular structure and processes

\section{Introduction}
\label{intro}

Reversible chemical modifications of proteins are well-known to play a pivotal role in the dynamics of the biochemical networks which allow a cell to convey and translate information from environmental signals to processes such as cell activation, proliferation and death \cite{dong,hol}. Moreover, some important biochemical paths are known to behave as irreversible on-off switches \cite{guna,qian}, and this switch-like character has been associated to the chemical modification dynamics of proteins on multiple aminoacid residues or domains \cite{gd}.
Multisite phosphorylation is the foremost example of protein modification because it  shows up ubiquitously in many important biochemical paths, but in addition there are several other multisite modification mechanisms, like acetylation, methylation, etc., which act at all levels in the biochemical control networks (see \cite{yang} for a recent review). Since multisite protein modification (MPM) is present in all eukariotes (yeasts, plants, and animal cells), it appears to be an evolutionary conserved mechanism that regulates biochemical thresholds and switching mechanisms. 

These considerations indicate that MPM is an essential component of many biochemical networks. However biochemical networks are complex entities where many processes are intertwined with one another and seem to be unapproachable with analytical tools: for this reason networks that incorporate MPM have been studied numerically in an effort to understand the role of MPM itself (see, e.g., \cite{ly}).

Here we attack the problem from a different standpoint: we assume that MPM steps have a common character and that they behave much like discrete components in an electronic circuit. Therefore we start by studying MPM in isolation: in this way it is possible to understand the role of MPM even without embedding it in a larger network, and we can produce a few analytical results before resorting to numerical methods. 

We have already discussed some of the nontrivial features of MPM in \cite{ourpap}: in the following section we briefly review the model and the concepts introduced in \cite{ourpap}. Section  \ref{diffeq}, where we derive the basic set of differential equations, and section \ref{eqval}, where we find the equilibrium concentrations, both review previous results and introduce the probabilistic interpretation that we have already used in \cite{ourpap}. Next we analyze the equations and find the system behavior for small deviations from the equilibrium values in sections \ref{relax} and \ref{eigen}. We use the results of section \ref{eigen} to synthesize noise spectra in section \ref{psd}. We obtain numerical results on the full  nonlinear system  in section \ref{num1} and discuss the general validity of the previous results in section \ref{open}. We include a downstream catalyzed Michaelis-Menten reaction in sections \ref{switch} and \ref{cyclecontrol}. We introduce a more realistic phosphorylation/dephosphorylation dynamics in section \ref{add} and use it to produce new numerical results in section \ref{cyclecontrol2}. Finally we draw our conclusions in section \ref{concl}.

\section{Stochastic analysis of multisite protein modification}
\label{review}

\begin{figure}
\centering
\includegraphics[width=3.2in]{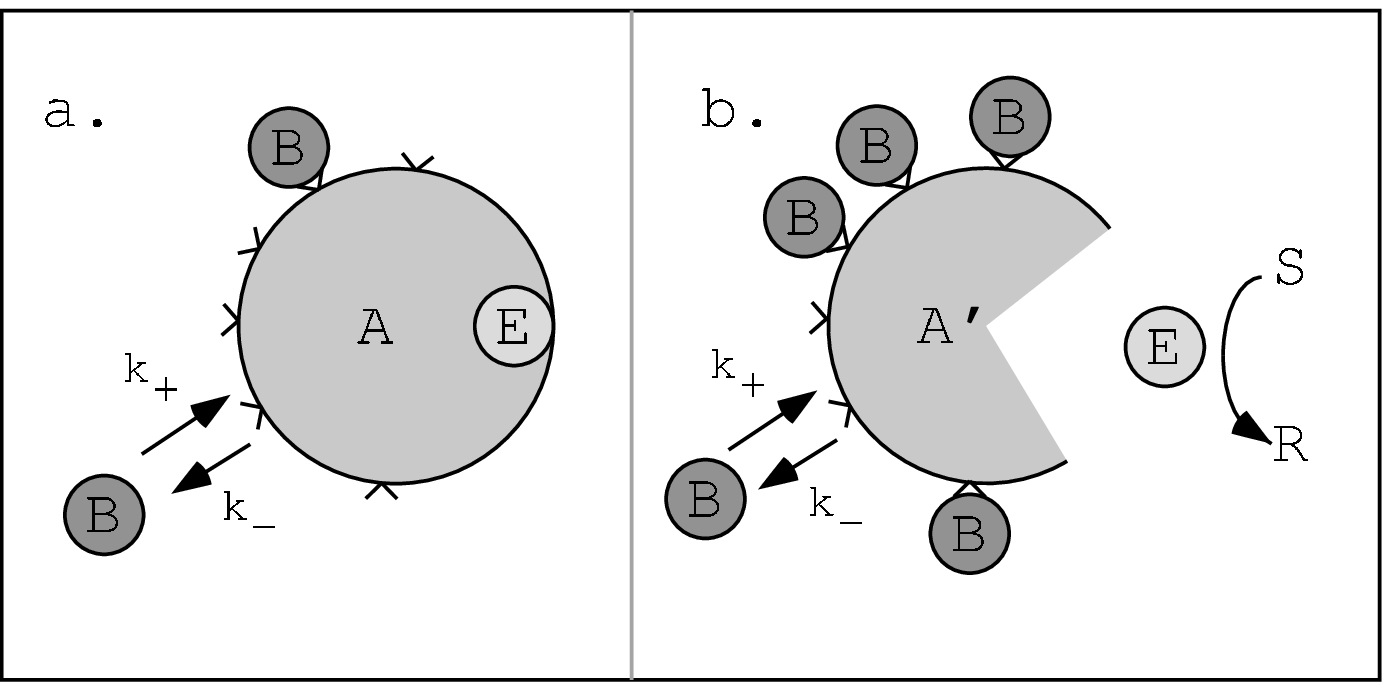}
\caption{\label{fig1} {\it a.} The figure shows schematically the system studied in this paper: the modification process is represented by molecules {\it B} that react with the sites  on {\it  A} with on-off rates  $k_+$, $k_-$. {\it b.} We also assume that when at least $n_{thr}$ sites out of the possible $N$ sites are occupied, molecule {\it A} is activated, and releases an enzyme {\it E} which catalyses a Michaelis-Menten reaction that converts a substrate {\it S} into a product {\it R}. In this paper we study the dynamics associated to the nonlinear system that describes this scheme. In this context we attach a probabilistic meaning to the results, we derive noise spectra, and study numerically multisite protein modification in conjunction with the downstream Michaelis-Menten step.}
\end{figure}

Here we summarize very briefly the results given in \cite{ourpap}: in that paper we introduce the reaction scheme shown in figure \ref{fig1}, which is very close to that considered by Monod, Wyman, and Changeux \cite{mwc},  where a chemical species $B$ can modify a number of sites on protein $A$. 
We also assume that the $N$ modifications  sites are all equivalent and that the modification dynamics for each site is independent from those of the other sites: this means that we consider the states $A_n$ with $n$ modified sites (see, e.g.,  the transition chain shown in figure 2 in \cite{ourpap}). Then, if the single chemical modification dynamics is fast with respect to the observation time we can forget the dynamics of the transition chain and even the chain itself and concentrate instead on the equilibrium probabilities. If the protein becomes activated when the number of modified sites is larger than a threshold value $n_{thr}$, then from the equilibrium probabilities $p_n$ of  the modified states $A_n$, we  show that the concentration of the activated form vs. $B$ has a sigmoid behavior with exponential tails, and this defines a very sharp biochemical threshold. The threshold turns out to be robust, i.e., it has a reduced sensitivity to parameter changes, which is further reduced as $N$ and $n_{thr}$ grow. Moreover, when we couple a downstream Michaelis-Menten process \cite{rubi,cb} we find that MPM can produce large delays, that once again depend on the number of modification sites $N$ and on $n_{thr}$, and can span several orders of magnitude, thereby providing a link between the fast time scale of molecular reactions and the slow pace of cell growth and proliferation.  We are also able to relate the model to the standard Hill phenomenology, which acquires a precise meaning in this context. And yet, the approach in \cite{ourpap} is incomplete because a real treatment of the dynamics is lacking, and to make further progress we must turn to a better dynamical description.

\section{Dynamical model of multisite protein modification}
\label{diffeq}

The model from which we start, and which we modify later to introduce an activation threshold, is a classic model in the theory of allosteric activity \cite{mwc,rubi}. The derivation of the equations is given, e.g., in \cite{rubi} section 2.4, and is based on the following simplifying assumptions:\begin{itemize}
\item all sites are equivalent;
\item the occupation of a given site is not influenced by the activity of nearby sites;
\item the number of modification sites is constant throughout the process (i.e., as the protein modification proceeds and possibly changes the protein shape, no new sites are added nor any existing sites are removed);
\item the behavior of the protein depends only on the total number of occupied sites, so that the state alone $A_n$ actually characterizes the protein activity;
\item we assume that the probability of multiple modification events is negligible, and therefore we consider only transitions to neighboring states (i.e., there are no transitions from $A_n$ to $A_{n+\Delta n}$ with $|\Delta n| > 1$).
\item the on-off rates $k_+$ and $k_-$ remain fixed and do not depend on modification-induced changes.
\end{itemize}

From these assumptions one finds the differential system 
\begin{eqnarray}
\nonumber
\frac{d[A_{0}]}{dt} & = & -N k_{+} [A_{0}] [B] + k_{-} [A_{1}]\\
\nonumber
&&\ldots\\
\nonumber
\frac{d[A_{n}]}{dt} & = & -n k_{-} [A_{n}] -(N-n) k_{+} [A_{n}] [B] +(N-n+1) k_{+} [A_{n-1}] [B] \\
\label{gensys}
&& + (n+1) k_{-} [A_{n+1}]\\
\nonumber
&&\ldots\\
\nonumber
\frac{d[A_{N}]}{dt} & = & -N k_{-} [A_{N}] + k_{+} [A_{N-1}] [B] 
\end{eqnarray}
Summing all the equations we find that A is conserved, and introducing the constant value $[A]_0$ (total concentration of $A$, which includes both the unmodified and the modified forms of $A$) we write the conservation equation 
\begin{equation}
\label{consA}
\sum_{n=0}^{N} [A_{n}] = [A]_0
\end{equation}

For the moment we also assume that total quantity of $B$ remains fixed, so that at any time the following conservation equation must hold as well
\begin{equation}
\label{Bcons}
\sum_{n=1}^{N} n[A_n] + [B] = [B]_0
\end{equation}
where the square brackets denote the molar concentrations, $[B]$ is the concentration of the {\em free} molecules of $B$,  and $[B]_0$ is the total concentration of $B$ assuming that all $B$'s are detached. 

We wish to stress that the nonlinear dynamical system described in this section is still highly idealized, it is chemically closed and in thermal equilibrium, and is not actually useful until it is coupled with the cellular environment: we accomplish this in section \ref{switch}, where we relax condition (\ref{Bcons}) (for a discussion of the features of closed and open biochemical systems see, e.g., the introduction of the review paper \cite{stu}).

\section{Equilibrium values}
\label{eqval}

As we mentioned above, the differential system (\ref{gensys}) is in the textbooks, and the equilibrium solution is well-known \cite{rubi}: in this section we review the basic results and recast the equilibrium solution in a suitable form. We introduce the auxiliary variables $p_n = [A_n]/[A]_0$, $r = (k_+/k_-)[B]$ and $b = [B]/[A]_0$, and the reduced parameter $s=(k_+/k_-)[A]_0$, so that $r=sb$. The $p_n$'s can be reinterpreted as the probabilities of finding a protein molecule with $n$ modified sites: with this probabilistic interpretation the differential system becomes the master equation for the $p_n$'s. We also introduce the corresponding barred quantities $\bar{p}_n$, $\bar{r}$ and $\bar{b}$, which denote the equilibrium values, and if we assume in addition that the underlying stochastic process (the chain of individual modification events) is ergodic, then in the long-time limit these probabilities also give the fraction of residence time in each modified state.
If the system is taken at equilibrium then the derivatives vanish and the differential system reduces to  a set of algebraic equations. The solution of the system is \cite{rubi}: 
\begin{equation}
\label{gensol}
\bar{p}_n = \binom{N}{n} \frac{\bar{r}^n}{(1+\bar{r})^N}
\end{equation}
Here $\bar{r}$ is still undefined, and we need yet another equation:  we take the conservation equation (\ref{Bcons}), which can be rewritten as
\begin{equation}
\label{gencons}
\sum_{n=1}^{N} n \bar{p}_n + \bar{b} = b_0
\end{equation}
where $b_0 = [B]_0/[A]_0$, then, substituting the solution (\ref{gensol}) in (\ref{gencons}) it is easy to show the new conservation condition
\begin{equation}
\label{bsol}
\frac{N\bar{r}}{1+\bar{r}} + \bar{b} = b_0
\end{equation}
and then we find a quadratic equation 
from which we get eventually the equilibrium concentration 
\begin{equation}
\label{genb}
\bar{b} = \frac{1}{2s}\left[ -(Ns+1-sb_0) + \sqrt{(Ns+1-sb_0)^2+4sb_0} \right]
\end{equation}
(the solution with the minus sign before the square root is unacceptable because it gives a negative concentration). The value (\ref{genb}) can be translated back to the usual notation so that the equilibrium concentration writes
\begin{eqnarray}
\nonumber
[B]_{eq} &=& \frac{1}{2}\left( -\left\{N[A]_0+(k_-/k_+)-[B]_0\right\} \right.\\
\label{genb2}
&& \left.+ \sqrt{\left\{N[A]_0+(k_-/k_+)-[B]_0\right\}^2+4(k_-/k_+)[B]_0} \right)
\end{eqnarray}
and the $\bar{r}$ value that is necessary to evaluate the $\bar{p}$'s is just $\bar{r} =(k_+/k_-) [B]_{eq}$.

At this point it is also interesting to notice that a directly observable quantity, the average occupation level, can be associated to the equilibrium values in a direct way: 
\begin{equation}
\langle n \rangle = \sum_{n=1}^{N} n \bar{p}_n = \frac{[B]_{0}-[B]_{eq}}{[A]_0}
\end{equation}
A straightforward calculation also yields  the variance of the fluctuations close to equilibrium
\begin{equation}
\operatorname{var}{n}  = \frac{N \bar{r}}{(1+\bar{r})^2}
\end{equation}

\begin{figure}
\centering
\includegraphics[width=3.2in]{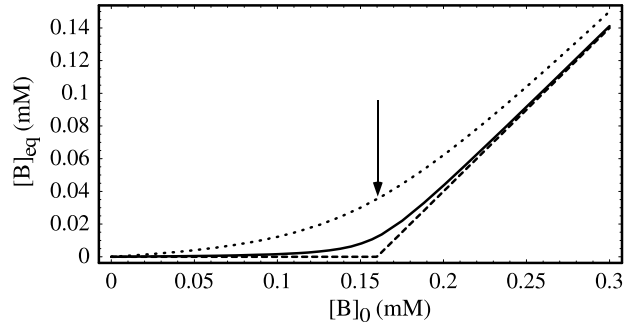}
\caption{\label{fig2} Equilibrium concentration $[B]_{eq}$ vs. $[B]_{0}$ for the example discussed in section \ref{eqval} (solid line), for a case with the same parameters but with a higher ratio $k_+/k_- =  10^{10}$ M (dashed line), and for a case with the same parameters but with a lower ratio $k_+/k_- =  10^{5}$ M (dotted line). The arrow marks the position of the threshold of saturation $[B]_{0} = N [A]_0$. }
\end{figure}

\begin{figure}
\centering
\includegraphics[width=3.2in]{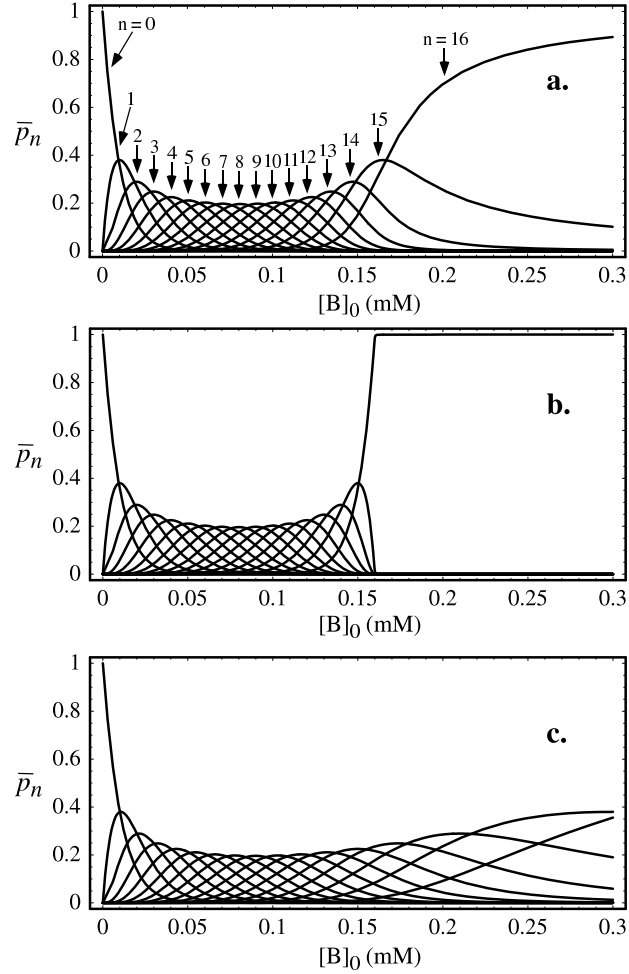}
\caption{\label{fig3} {\bf a}). Equilibrium probabilities $\bar{p}_n$ vs. $[B]_{0}$ for the example discussed in section \ref{eqval}. The curves for different $\bar{p}_n$'s are labeled accordingly. {\bf b}). Equilibrium probabilities for the same parameters but with a higher ratio $k_+/k_- =  10^{10}$ M: notice that in this case $\bar{p}_{16}$ reaches saturation as soon as $[B]_0$ reaches the threshold level $[B]_0 = 0.16$ mM. {\bf c}). Equilibrium probabilities for the same parameters but with a lower ratio $k_+/k_- =  10^{5}$ M: in this case saturation is approached much more slowly. }
\end{figure}

\begin{figure}
\centering
\includegraphics[width=3.2in]{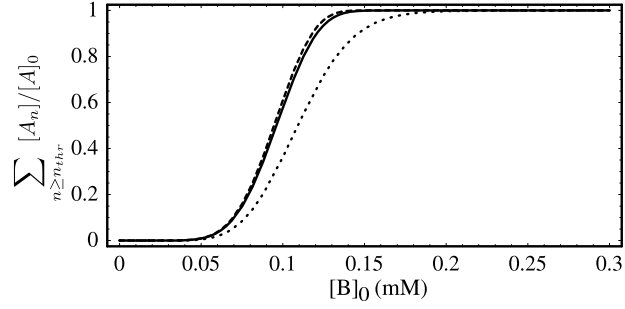}
\caption{\label{fig4} Relative concentration $\sum_{n\ge n_{thr}}[A_n]/[A]_0$ of the modified $A_n$'s that are above a threshold value $n_{thr}$ ($n_{thr}=10$ in this example). Solid line: parameter values as in the example discussed in section \ref{eqval}; dashed line: same parameters but with a higher ratio $k_+/k_- =  10^{10}$ M; dotted line (lowest curve): same parameters but with a lower ratio $k_+/k_- =  10^{5}$ M. }
\end{figure}

\begin{figure}
\centering
\includegraphics[width=3.2in]{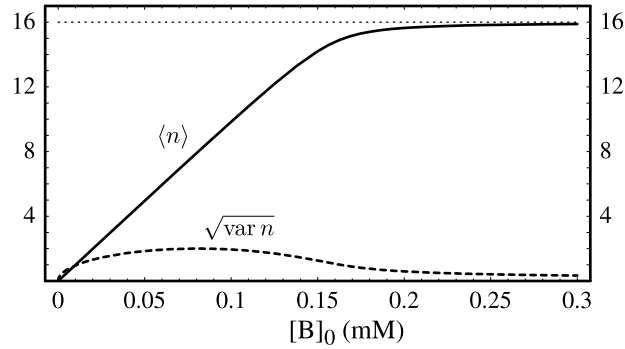}
\caption{\label{fig5} Solid line: average number of occupied sites $\langle n \rangle$ vs. $[B]_{0}$ for the example discussed in section \ref{eqval}; dashed line: standard deviation of the number of modified sites $\sqrt{\operatorname{var}n}$. The thin dotted line shows the saturation value, $n = N = 16$ in this case. Notice that the average $\langle n \rangle$ grows linearly until saturation, while the standard deviation is almost always much smaller than the average, and decreases for high values of the concentration $[B]_{0}$ .}
\end{figure}

\begin{figure}
\centering
\includegraphics[width=3.2in]{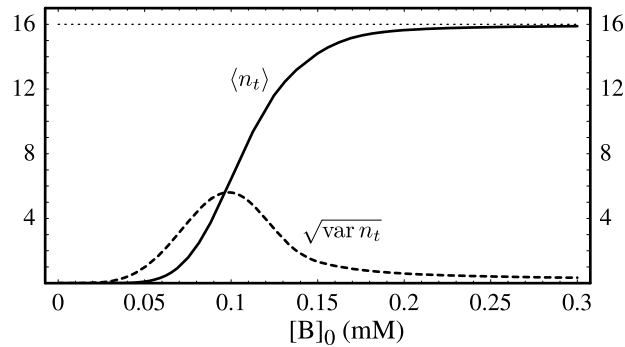}
\caption{\label{fig6} Statistics of the number of occupied sites $n_t$ above a given threshold value $n_{thr}$ ($n_{thr}=10$ in this example). Solid line: average number of occupied sites $\langle n_t \rangle$ vs. $[B]_{0}$; dashed line: standard deviation $\sqrt{\operatorname{var}n_t}$. The thin dotted line shows the saturation value, $n = N = 16$ in this case.}
\end{figure}

The equilibrium values $[A_n]_{eq}$ and $[B]_{eq}$ depend on the total concentrations $[A]_0$, $[B]_0$, and on the on-off ratio $k_+/k_-$. Whenever $(k_+/k_-)$ is large, the equilibrium concentration (\ref{genb2}) can be approximated as follows
\begin{equation}
\label{genb3}
[B]_{eq}\approx \frac{1}{2}\left[   |N[A]_0-[B]_0| -\left(N[A]_0-[B]_0\right) \right]
\end{equation}
and we see that there is a breakpoint at $[B]_0=N[A]_0$, which is the threshold of saturation. This behavior is illustrated graphically in figure \ref{fig2} which shows  $[B]_{eq}$ vs. $[B]_{0}$, and  where we have set $[A]_0 = 10 \mu$M, which corresponds roughly to the concentration of the most abundant proteins in a cell, $k_+/k_-=10^{6}$ M, which is a common value for the on-off ratio (see \cite{rubi}, p. 56),  and we have taken $N = 16$ which is the same as the number of putative phosphorylation sites for the Rb protein \cite{ez,sev}.  Figure \ref{fig3} shows the corresponding probabilities $\bar{p}_n$ vs. $[B]_{0}$, and figure \ref{fig4} shows the relative concentration $\sum_{n\ge n_{thr}}[A_n]/[A]_0$ where $n_{thr}$ is a threshold number of modification sites as in \cite{ourpap} (in this example $n_{thr} = 10$): although the probabilities in figure \ref{fig3} change considerably when the ratio $k_+/k_-$ is varied, the relative concentration of the modified A's above threshold is remarkably stable with respect to changes of the $k_+/k_-$ ratio. Notice that the inverse ratio $k_-/k_+=10^{-6}$ M is close both to $[A]_0$ and $[B]_{0}$, and for this reason figures \ref{fig2} and \ref{fig3} also include the behavior of $[B]_{eq}$ and of the $p_n$'s for both smaller and larger values of the on-off ratio. We note that for large values of  $k_+/k_-$, i.e. in the case in which the detachment reaction is negligible with respect to site modification, the curves are more kinky and change sharply after reaching the saturation threshold. 
Figure \ref{fig5} shows the modification level $\langle n \rangle$ and the standard deviation $\sqrt{\operatorname{var} n}$. Notice that $\langle n \rangle$ grows nearly linearly until the saturation value is reached, at $[B]_0=N[A]_0=0.16$ mM, and that the standard deviation is usually much smaller than the average modification level. Similarly, figure \ref{fig6} shows the average number of modified sites in the $A_n$'s that are above threshold, i.e., $\langle n_t \rangle = \sum_{n\ge n_{thr}} n \bar{p}_n$, and the corresponding standard deviation $\sqrt{\operatorname{var} n_t}$ of the number of sites above threshold: the standard deviation $\sqrt{\operatorname{var} n_t}$ is the largest in the vicinity of the threshold.

\section{Linearized dynamics}
\label{relax}

We have already noted in section \ref{diffeq} that the system is closed, and therefore -- even though the equations are nonlinear -- we can state on very general grounds that it must be stable as well \cite{stu,shear}.
In the previous section we have examined the equilibrium values, but in all possible biological settings it is very important to consider the dynamical behavior of the concentrations $[A_n]$ and of the modification level $\langle n \rangle$ as well, and we turn again to the original differential system (\ref{gensys}), which we rewrite here using the reduced variables: 
\begin{eqnarray}
\nonumber
&&\frac{dp_0}{dt} = k_{-} \left\{-Nrp_0 + p_1\right\} \\
\nonumber
&&\ldots\\
\label{gensys2}
&&\frac{dp_n}{dt} = k_{-} \left\{-n p_n -(N-n) r p_n +(N-n+1) r p_{n-1} + (n+1) p_{n+1} \right\} \\
\nonumber
&&\ldots\\
\nonumber
&&\frac{dp_N}{dt} = k_{-} \left\{-N p_N+ r p_{N-1} \right\}
\end{eqnarray}
where we wish to stress that, in addition to the $[A_n]$'s, also $[B]$ and therefore also $r$ are time-dependent quantities. If we introduce the deviations from the equilibrium values $\Delta p_n = p_n - \bar{p}_n$ and recall that $ b = b_0-\sum_{n=1}^{N} n p_n = \bar{b} - \sum_{n=1}^{N} n \Delta p_n$, we can linearize the system (\ref{gensys2}) for small deviations: 
\begin{eqnarray}
\nonumber
\frac{d\Delta p_0}{dt} &=& k_{-} \left\{-Ns[\bar{b}\Delta p_0-\bar{p}_0 \sum_{m=1}^N m \Delta p_m] + \Delta p_1\right\} \\
\nonumber
&&\ldots\\
\nonumber
\frac{d\Delta p_n}{dt} &=& k_{-} \left\{-n \Delta p_n -(N-n) s [ \bar{b}\Delta p_n - \bar{p}_n \sum_{m=1}^N m \Delta p_m] \right.\\
\label{linsys}
&&\left. + (N-n+1)s [\bar{b} \Delta p_{n-1} - \bar{p}_{n-1} \sum_{m=1}^N m \Delta p_m]+ (n+1)\Delta p_{n+1} \right\} \\
\nonumber
&&\ldots\\
\nonumber
\frac{d\Delta p_N}{dt} &=& k_{-} \left\{-N \Delta p_N+ s [\bar{b} \Delta p_{N-1} - \bar{p}_{N-1} \sum_{m=1}^N m \Delta p_m]\right\}
\end{eqnarray}

In most cases even the linearized dynamics can only be studied numerically \cite{stu}, however here we are able to derive the eigenvalues of the system matrix \cite{bn}, and thus to evaluate all the characteristic time scales of the system.

\section{Eigenvalues}
\label{eigen}

The generic linearized equation in the differential system (\ref{linsys}) is 
\begin{eqnarray}
\nonumber
\frac{1}{k_-} \frac{d\Delta p_n}{dt} & = & \left\{ (N-n+1) \bar{r}\Delta p_{n-1} - [n+(N-n)\bar{r}]\Delta p_n + (n+1) \Delta p_{n+1} \right\} \\
&&+ s\left[ (N-n) \bar{p}_n - (N-n+1) \bar{p}_{n-1} \right] \sum_{m=1}^N m\Delta p_m
\end{eqnarray}
If we temporarily drop the term proportional to the sum $\sum_{m=1}^N m\Delta p_m$, and set $k_- =1$, we are left with a system matrix that does not have a definite symmetry, but shows a remarkably ordered structure: 
\begin{equation}
\label{mat0}
\mathbf{A}_N^{(r)} = \left(\begin{array}{ccccc}
-N\bar{r} & 1 & 0 & 0 & \cdots \\
N\bar{r} & -1-(N-1)\bar{r} & 2 & 0 & \cdots \\
0 & (N-1)\bar{r} & -2-(N-2)\bar{r} & 3 & \cdots \\
0 & 0 & (N-2)\bar{r} & 0 & \cdots \\
\vdots & \vdots & \vdots & \vdots & \ddots\end{array}\right)
\end{equation}
In order to find the eigenvalues of $\mathbf{A}_N^{(r)}$ we introduce the matrix $\mathbf{U}_N$: 
\begin{equation}
\left\{ \mathbf{U}_N \right\}_{j,k} = 
\left\{
\begin{array}{cc}
0 & j<k \\
\binom{N-k}{j-k} & j \ge k
\end{array}
\right.
\end{equation}
and its inverse
\begin{equation}
 \left\{ \mathbf{U}_N^{-1} \right\}_{j,k} = 
\left\{
\begin{array}{cc}
0 & j<k \\
(-1)^{j+k} \binom{N-k}{j-k} & j \ge k
\end{array}
\right.
\end{equation}
Both $\mathbf{U}_N$ and its inverse are lower triangular matrices, with a single degenerate eigenvalue $\lambda = 1$.  
We use $\mathbf{U}_N$ to perform a basis change, and we obtain
\begin{equation}
\left\{ \mathbf{U}_N \mathbf{A}_N^{(r)} \mathbf{U}_N^{-1} \right\}_{j,k} = k\delta_{j,k-1} - \left[ (N-k) + (N-k)\bar{r} \right]\delta_{j,k}
\end{equation}
which shows that the eigenvalues of $\mathbf{A}_N^{(r)}$ are $\lambda_k^{(r)} = - \left[ (N-k) + (N-k)\bar{r} \right]$ ($0 \le k \le N$). 

The actual system matrix is $\mathbf{A} = \mathbf{A}_N^{(r)}+\mathbf{A}_N^{(s)}$, where $\mathbf{A}_N^{(s)}$ corresponds to the dropped term proportional to $\sum_{m=1}^N m\Delta p_m$. The elements of the $\mathbf{A}_N^{(s)}$ matrix are: 
\begin{equation*}
\begin{array}{lclr}
\left\{ \mathbf{A}_N^{(s)} \right\}_{0,k} & = & Nks\bar{p}_0 &   \\
\left.\left\{ \mathbf{A}_N^{(s)} \right\}_{j,k}\right|_{1 \le j < N} & = & k\left[ (N-j)s\bar{p}_j - (N-j+1)s\bar{p}_{j-1}\right] &  \\
\left\{ \mathbf{A}_N^{(s)} \right\}_{N,k} & = & -ks\bar{p}_{N-1} &  
\end{array}
\end{equation*}

In the new basis we find 
\begin{equation*}
\begin{array}{lclr}
\left\{ \mathbf{U}_N \mathbf{A}_N^{(s)} \mathbf{U}_N^{-1} \right\}_{0,k} & = & Ns\bar{p}_0\left(N\delta_{N,k}-\delta_{N-1,k}\right) &   \\
\left.\left\{ \mathbf{U}_N \mathbf{A}_N^{(s)} \mathbf{U}_N^{-1} \right\}_{j,k}\right|_{1 \le j < N} & = & s\left\{ \binom{N-1}{j}N\bar{p}_0+\sum_{l=1}^{j-1}\binom{N-l-1}{j-l}(N-l)\bar{p}_l \right. & \\
&& \left.+(N-j)\bar{p}_j \right\} \left(N\delta_{N,k}-\delta_{N-1,k}\right) &  \\
\left\{ \mathbf{U}_N \mathbf{A}_N^{(s)} \mathbf{U}_N^{-1} \right\}_{N,k} & = & 0 &  
\end{array}
\end{equation*}
and from this result we see that the eigenvalues of the complete system matrix are 
\begin{equation}
\begin{array}{lclr}
\left.\lambda_{j}\right|_{0 \le j < N-1} & = & -k_-\left[ (N-j)+(N-j)s\bar{b}\right] &  \\
\lambda_{N-1} & = & -k_-\left[ 1+s\bar{b}+s\sum_{l=0}^{N-1} (N-l) \bar{p}_l \right] &   \\
\lambda_N & = & 0 &  
\end{array}
\end{equation}
(where the off rate $k_-$ has been restored to its original value).
We have seen earlier that the conservation equation (\ref{consA}), which translates into the normalization condition for the probabilities $p_n$, is automatically satisfied by the differential system (\ref{gensys}), and  it can be readily verified that the linearized system (\ref{linsys}) preserves this condition: this produces the 0 eigenvalue. 

Figure \ref{fig7} shows plots of the the eigenvalues vs. $[B]_0$ for the example discussed in section \ref{eqval} (i.e. with $N=16$): apart from the even spacing of the eigenvalues $\lambda_0$ to $\lambda_{14}$, one important feature of this plot is the cross-over behavior of the $\lambda_{15}$ eigenvalue. Using the equilibrium concentration $\bar{b}$ given by equation (\ref{genb}) and the cross-over condition $\lambda_{N-2} = \lambda_{N-1}$, after some straightforward algebra one finds the cross-over concentration
\begin{equation}
b_{0,{\mathrm crossover}} = \frac{Ns -1}{s}
\end{equation}
which approximates to $b_{0,{\mathrm crossover}} \approx N$, i.e. $[B]_0 \approx N [A]_0$ for $s \gg 1$ (i.e., in this case the cross-over value corresponds to the saturation threshold).
Notice also that for fixed concentrations of A and B, and fixed on-off ratio $k_+/k_-$, the system moves to higher frequencies (shorter reaction times) as $N$ grows, i.e., fluctuations are more effectively damped-off for higher $N$'s.

\section{Synthetic noise spectra}
\label{psd}

Biochemical reactions where only few molecules are involved are affected by molecular noise: this noise often has a deep biochemical meaning \cite{rao}, and is most often studied by Monte Carlo simulation \cite{rao,gill}. However, when the eigenvalues of the linearized system are known, as in the present case, it is possible to synthesize directly the noise spectra \cite{simpson}. The eigenvalues  determine the spectral density of the occupation level of the molecular population: because of the random (Poisson) character of the individual molecular events the spectral density of the concentrations in the dynamical system, and in particular the spectral density $S(f)$ of the modification level $n$, can be derived from the incoherent superposition of the spectral densities of the individual relaxation processes associated to each eigenvalue \cite{simpson,bernamont}
\begin{equation}
S(f) \propto \sum_{n=0,N-1} \frac{1}{\lambda_n^2+(2\pi f)^2}
\end{equation}

\begin{figure}
\centering
\includegraphics[width=3.2in]{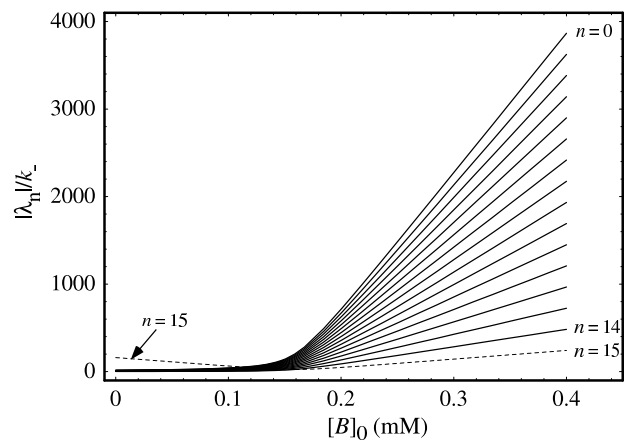}
\caption{\label{fig7} Eigenvalues $|\lambda_n|/k_-$ vs. $[B]_{0}$ for the example discussed in section \ref{eqval} (with $N=16$): the eigenvalues from $\lambda_0$ up to $\lambda_N = \lambda_{14}$ are evenly spaced, while the eigenvalue $\lambda_{15}$ is the highest for low $[B]_0$, crosses over the distribution of the other eigenvalues, and ends up as the lowest eigenvalue for high $[B]_0$. }
\end{figure}

We point out that the noise spectrum of the fluctuating modification level has a characteristic shape that depends on the concentration of $B$.  The eigenvalue distribution for the example discussed at the end of section \ref{eqval} and shown in figure \ref{fig7} indicates that at low concentration $[B]$ the spectral density is roughly the superposition of two simple relaxation processes, then close to the threshold the spectral density collapses to a simple relaxation process (and thus is characterized by a $1/f^2$ power-law region). Finally, above threshold the spectral density has low-frequency white noise region, an intermediate $1/f$ region, and a $1/f^2$ high-frequency behavior. The  $1/f$ power-law behavior is limited to the range determined by the minimum frequency $f_{min} = k_-(\mathop {\min }\limits_{0 \le n < N} \lambda_n)/2\pi$ and the maximum frequency $f_{max} = k_-(\mathop {\max }\limits_{0 \le n < N} \lambda_n)/2\pi$. 

\begin{figure}
\centering
\includegraphics[width=3.2in]{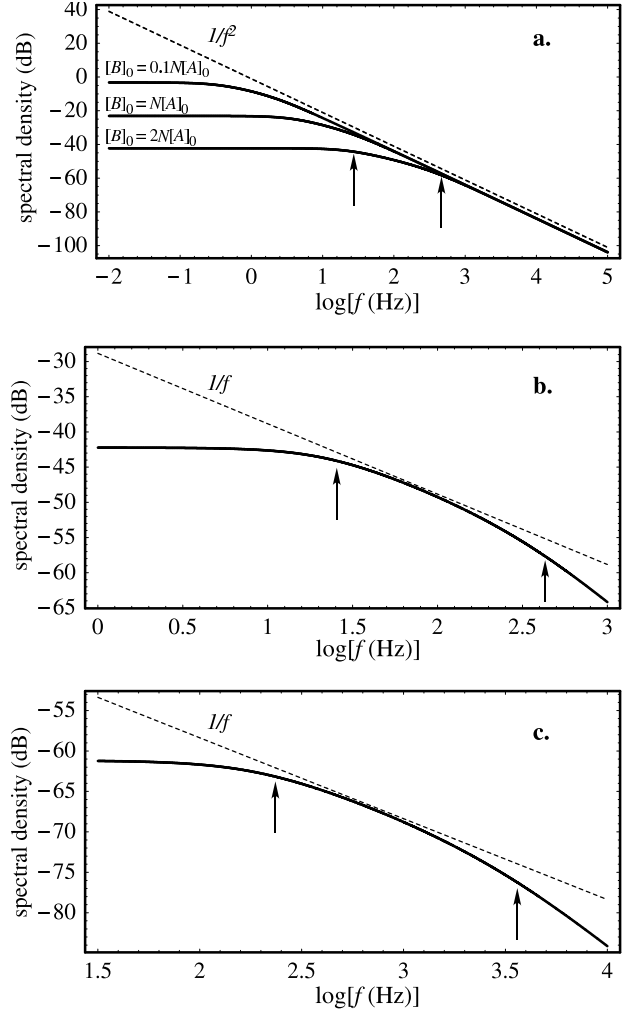}
\caption{\label{fig8} These figures display some synthetic spectra of the number of $n(t)$ of occupied sites, obtained from the eigenvalues $|\lambda_n|/k_-$ vs. $[B]_{0}$ for the example discussed in section \ref{eqval} (with $N=16$) and shown in figure \ref{fig7}. The amplitude scale is in arbitrary units and may change for different spectra. a) spectra obtained with $[B]_{0} = 0.1 N [A]_{0}$, $[B]_{0} =  N [A]_{0}$, and $[B]_{0} = 2 N [A]_{0}$ (solid lines) and an ideal $1/f^2$ noise spectrum (dotted line): the first two spectra have a low-frequency white noise plateau and a high-frequency $1/f^2$ noise tail. The $[B]_{0} = 2 N [A]_{0}$  case also shows a limited $1/f$ noise region, between the arrows (which mark the position of the extreme eigenvalues $\lambda_0$ and $\lambda_{15}$). b) Close-up of the $1/f$ noise region for the case $[B]_{0} = 2 N [A]_{0}$; here the dotted line is an ideal $1/f$ spectrum and the arrows mark the positions of the extreme eigenvalues. c)  A larger value $[B]_{0} = 10 N [A]_{0}$ moves the $1/f$ noise region to higher frequency; once again the dotted line is an ideal $1/f$ spectrum and the arrows mark the positions of the extreme eigenvalues.}
\end{figure}

All this is further illustrated in figure \ref{fig8} which shows some synthetic spectra obtained from the eigenvalues shown in figure \ref{fig7}.
Figure \ref{fig8} shows that the spectra for the example in section \ref{eqval} have a low-frequency white noise plateau and a high-frequency $1/f^2$ noise tail, however when $[B]_0 \ge N [A]_0$ there is also a small $1/f$ noise region which spans approximately one frequency decade just as discussed above. However, although the slope change is clear, the $1/f$ region is not well defined because of the closeness of the extreme eigenvalues $\lambda_0$ and $\lambda_{15}$. 

\section{Numerical solution of the differential system}
\label{num1}

\begin{figure}
\centering
\includegraphics[width=3.2in]{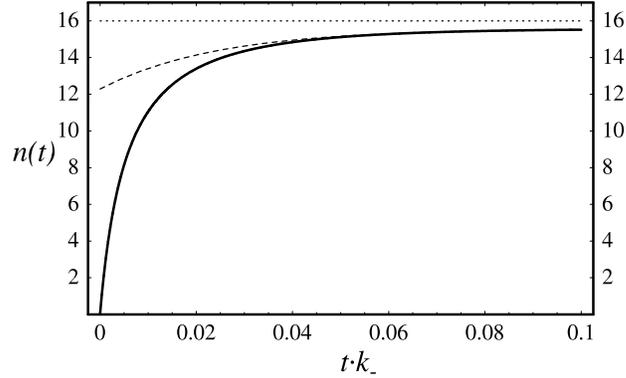}
\caption{\label{fig9} Behavior of $n(t)$ from the numerical integration of the differential system (\ref{gensys}) (solid curve) with the conditions specified in section \ref{num1}. The dashed curve is a fit of the tail for $t>0.2 s$ with the function $c_1 - c_2 \exp(-\lambda t)$: we find $\lambda = 41.5851/k_-$, which is very close to theoretical value of the maximum eigenvalue (i.e. largest absolute value: $|\lambda_{15}| = 41.5812/k_-$). The number of occupied sites asymptotically approaches the equilibrium value, here $\langle n \rangle \approx 15.5709$, which is slightly smaller than the saturation value $N =16$ (thin dotted line). }
\end{figure}

The full differential system (\ref{gensys}) may be solved numerically with standard integration methods: we remark that the (asymptotic) stability properties are the same as those of the linearized system, and are guaranteed in our case by a theorem due to Poincar\'e and Perron (see, e.g. \cite{bn}, pp. 161-163), and therefore we do no expect to find any remarkably new features in the numerical solutions.
We have integrated the differential system (\ref{gensys}) assuming the values at the end of section \ref{eqval}, i.e., $N=16$, $k_-=1$ s$^{-1}$, $k_+ = 10^6$ s$^{-1}$ M$^{-1}$, and with the initial conditions $[A_0]_{t=0} = [A]_0 = 10 \mu$M, $[B]_{t=0}=[B]_0=1.2 N [A]_0$, and $[A_n]_{t=0} = 0$ for $n>0$. 
Figure \ref{fig9} shows the behavior of the number of occupied sites $n$ vs. the dimensionless time variable $t\cdot k_-$: the fitting exponential for long times is also shown, and the corresponding time constant is in excellent agreement with the value computed in the previous section (i.e., the maximum nonzero eigenvalue).

\section{Opening up the system}
\label{open}

Up till now we have studied MSM in isolation, and -- apart from their utility in computing the noise spectra -- it is natural to wonder if the eigenvalues can also be useful to understand MSM when the biochemical system is open. Here we consider a straightforward modification, we assume that each equation in the differential system (\ref{gensys}) contains an additional term $-a_n [A_n]$ (we include a minus sign because this additional term is usually dissipative). With these additional terms the system matrix changes: $\mathbf{A}_N^{(r)} \rightarrow \mathbf{A}_N^{(r)} - \operatorname{diag}{n}(a_0, \ldots, a_N)$, where $ \operatorname{diag}{n}(a_0, \ldots, a_N)$ is a diagonal matrix with diagonal elements $a_0, \ldots, a_N$. Using the transformation matrix $\mathbf{U}_N$ and its inverse, it is easy to show that the eigenvalues transform as follows: $\lambda_n \rightarrow \lambda_n - a_n$. 
For this reason the previous calculation of the eigenvalues retains its value and can be used to estimate the behavior of the system even when it is no longer closed and thus when the principle of detailed balance no longer holds.

In the next section we turn to a more complex modification of the system, one which involves the coupling to a downstream reaction.

\section{Switched downstream Michaelis-Menten process}
\label{switch}

In this section we consider the following set of coupled reactions
\[\left\{ A_{n}+B \underset{k_+}{\overset{k_-}{\leftrightarrows}} A_{n+1} \right\}_{n=0,\ldots,N-1} \;\;\;\;\; \left\{ A_{n} \underset{k_{E+}}{\overset{k_{E-}}{\leftrightarrows}} A^{\prime}_{n} + E \right\}_{n=n_{thr},\ldots,N},\] 
\[E + S \underset{k_1}{\overset{k_2}{\leftrightarrows}} ES \overset{k_3}{\longrightarrow} E + R.  \] 
where $n_{thr}$ is the threshold modification level mentioned above, which corresponds to the onset of release of the secondary enzyme $E$: the modified species $A_n$ changes to $A^\prime_n$ and releases $E$. The last reaction is a Michaelis-Menten step catalyzed by $E$ which acts on a substrate $S$ and produces $R$. We also make the additional simplifying hypothesis that $A^\prime_n$ can no longer take part to the modification chain, and can reenter the chain only after reabsorbing $E$. 
With these assumptions the differential system (\ref{gensys}) becomes: 
\begin{eqnarray}
\nonumber
\frac{d[A_{0}]}{dt} & = & -N k_{+} [A_{0}] [B] + k_{-} [A_{1}]\\
\nonumber
&&\ldots\\
\nonumber
\left. \frac{d[A_{n}]}{dt}\right|_{n < n_{thr}} & = & -n k_{-} [A_{n}] -(N-n) k_{+} [A_{n}] [B] +(N-n+1) k_{+} [A_{n-1}] [B] \\
\nonumber
&& + (n+1) k_{-} [A_{n+1}] \\
\nonumber
&&\ldots\\
\nonumber
\left. \frac{d[A_{n}]}{dt}\right|_{n \ge n_{thr}} & = & -n k_{-} [A_{n}] -(N-n) k_{+} [A_{n}] [B] +(N-n+1) k_{+} [A_{n-1}] [B] \\
\nonumber
&& + (n+1) k_{-} [A_{n+1}] - k_{E+}[A_n] + k_{E-}[A_n^\prime][E] \\
\label{MMsys}
&&\ldots\\
\nonumber
\frac{d[A_{N}]}{dt} & = & -N k_{-} [A_{N}] + k_{+} [A_{N-1}] [B] - k_{E+}[A_N] + k_{E-}[A_N^\prime][E]
\end{eqnarray}
In addition to the equations for the $[A_n]$'s we must also add the equations for the $[A_n^\prime]$'s and for the enzyme E: 
\begin{eqnarray}
\nonumber
\left.\frac{d[A_{n}^\prime]}{dt}\right|_{n \ge n_{thr}} & = & k_{E+}[A_n] - k_{E-}[A_n^\prime][E]\\
\frac{d[E]}{dt} & = & \sum_{n=n_{thr}}^N k_{E+} [A_n] - \sum_{n=n_{thr}}^N k_{E-} [A_n^\prime][E] + \frac{d[E]_{MM}}{dt}
\end{eqnarray}
where the derivative $d[E]_{MM}/dt$ is the contribution of the Michaelis-Menten step: 
\begin{eqnarray}
\frac{d[R]}{dt} & = & k_3 [ES]\\
\frac{d[S]}{dt} & = & -k_1[E][S]+k_2[ES] + s(t)\\
\frac{d[E]_{MM}}{dt} & = & -k_1[E][S] + k_2[ES] + k_3 [ES]\\
\frac{d[ES]}{dt} & = & k_1[E][S] - k_2[ES] - k_3 [ES] = -\frac{d[E]_{MM}}{dt}
\end{eqnarray}
(the term $s(t)$ is the rate with which the substrate S is replenished). 
As before, these equations must be complemented by a conservation equation for $[B]$ which now writes:
\begin{equation}
\label{Bconsnew}
\sum_{n=1}^N n[A_n] + \sum_{n=n_{thr}}^N n [A_n^\prime] + [B] = [B]_0 
\end{equation}

\begin{figure}
\centering
\includegraphics[width=3.2in]{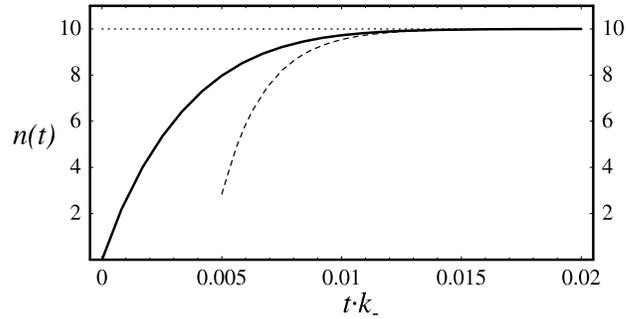}
\caption{\label{fig10} Behavior of $n(t)$ from the numerical integration of the differential system in section \ref{switch}, without the inclusion of the Michaelis-Menten reaction (i.e., only the equations for the $A_n$'s and for the $A'_n$'s are included) and with the parameters specified in section \ref{cyclecontrol}. The dotted curve is the fit of the tail for $t>0.022 s$ with the function $c_1 - c_2 \exp(-\lambda t)$: we find $\lambda \approx 550$ Hz, which is much larger than the decay constant found in the example shown in figure \ref{fig9}. The number of occupied sites quickly approaches the threshold value $n_{thr}=10$. }
\end{figure}

We have studied numerically the new modified system: we have taken the same conditions as in section \ref{num1} and in addition we have set $n_{thr}=10$ and $[A'_n]_{t=0} = 0$. 
Figure \ref{fig10} shows the behavior of the number of occupied sites $n$ vs. the dimensionless time variable $tk_-$: the fitting exponential for long times is also shown: now the decay constant is much larger than that found integration shown in figure \ref{fig9}, i.e., when $[B]$ is above the critical value, the dynamical system reacts very quickly to environmental changes. The inclusion of the Michaelis-Menten part does not change this behavior, and the approach to equilibrium is faster.

\section{Cell-cycle control}
\label{cyclecontrol}

The modified system with the downstream Michaelis-Menten step is reminiscent of the way the cyclin-CDK complex phosphorylates the Rb protein which then releases the transcription factor E2F \cite{mbc}, which is an important step in the cell-cycle, since it leads to the so-called {\em G1-S checkpoint}  \cite{ez,sev}. Even when we leave aside the enormous complexity of cell-cycle regulation as a whole and concentrate on an important detail like Rb protein activity, we are still left with a very complicated pattern of biochemical reactions (see, e.g., the figure depicting the Rb network in \cite{jdg}). However the model can be further simplified taking only some essential elements from this network (see, e.g., figure 8 in \cite{rs}, which is a good introduction for physicists, see also \cite{sha}), and in particular we assume that: 
\begin{itemize}
\item the cyclin is destroyed during the cell cycle and is synthesized again after mitosis;
\item we neglect the difference between cyclin D and cyclin E;
\item there is plenty of ATP in the environment (cytosol) and we assume that the total concentration of phosphoryl groups, i.e., both those in the ATP bound to the cyclin-CDK complex and the phosphoryl groups bound to the Rb protein, rises roughly linearly as cyclin is produced during the early G1 phase \cite{pines} (these phosphoryl groups effectively represents $B$); 
\end{itemize}
To simulate this process we have integrated numerically the set of equations  (\ref{MMsys})-(\ref{Bconsnew}) with the condition 
\begin{equation}
[B]_0 = B_r t
\end{equation}
where $B_r$ is the production rate of available phosphoryl groups and we assume that this production rate is small in comparison to the natural relaxation rates of the system (i.e., the previously calculated eigenvalues) so that the considerations of the previous sections which assume a constant $[B]_0$, i.e.,  $B_r=0$, still apply. Notice also that the system is no longer closed and that the principle of detailed balance does not hold in this modified situation.

We keep the conditions that we have already used in the previous numerical integrations, and in addition we take: $k_{E+}=10^7$ s$^{-1}$, $k_{E-}=1$ s$^{-1}$ M$^{-1}$, $k_1=10^7$ s$^{-1}$ M$^{-1}$, $k_2=k_3=10^3$  s$^{-1}$,  the initial conditions $[E]_{t=0}=[ES]_{t=0}=[S]_{t=0}=[R]_{t=0}=0$ (these values are in the range of values for Michaelis-Menten process commonly found in cells, see, e.g., \cite{cb}, p. 39).

\begin{figure}
\centering
\includegraphics[width=3.2in]{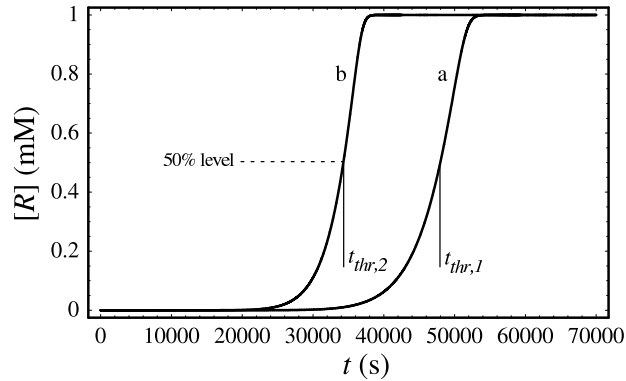}
\caption{\label{fig11} This figure shows the concentration $[R]$ of the product of the downstream Michaelis-Menten reaction of section \ref{cyclecontrol}, when we assume that the concentration of the enzyme $B$ that modifies the substrate $A$ grows linearly in time. Curve {\bf a} has been obtained with the parameters given in section \ref{switch} and with the $B$ production rate $B_r = N [A]_0/T$, while curve {\bf b} has been obtained with the higher rate $B_r = 1.5 N [A]_0/T$. We define a threshold level that is 50\% of the saturation level of $R$ and we find the corresponding times $t_1$ and $t_2$. The large (50\%) change in production rate leads to a smaller change in threshold time. }
\end{figure}

Figure \ref{fig11} shows $[R]$ for two values of the production rate ($B_r =N [A]_0/T$ and $B_r = 1.5 N [A]_0/T$,  where $T = 10^6$ s): these curves are very similar to those that we had obtained in \cite{ourpap} using a simple approximation. Although the production rates have a 50\% difference the curves are quite close, and this suggests that multisite phosphorylation helps making the system robust with respect to changes in production rate. 

\begin{figure}
\centering
\includegraphics[width=3.2in]{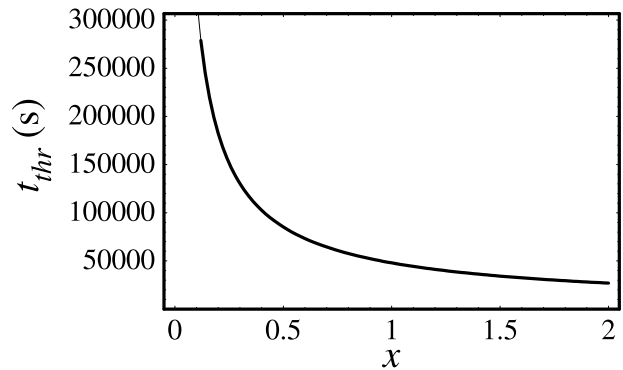}
\caption{\label{fig12} Threshold time (defined in figure \ref{fig11}) vs. $B$ production rate, from the numerical integration of section \ref{cyclecontrol}. The curve is very well fit by the function  $t_{thr}(x) = a + (b/x)^\alpha$, where $x= B_r/(N [A]_0/T)$ is the relative production rate, with $a = 81.7$ s, $b=4.32 \cdot 10^5$ s$^{1/\alpha}$, and $\alpha = 0.83$ (fit and numerical results are so close that they are indistinguishable from each other in this figure).  }
\end{figure}

If we use the point at half maximum as representative of the thresholding behavior, we can study the  threshold position vs. the synthesis rate $B_r$. This is shown by the curve in figure \ref{fig12}: the curve is well fit by a function with a power-law component
\begin{equation}
t_{thr}(x) = a + (b/x)^\alpha
\end{equation}
where $x= B_r/(N [A]_0/T)$ is the relative production rate, so that the relative change of $t_{thr}$ is 
\begin{equation}
\frac{\Delta t_{thr}}{t_{thr}} = \frac{\alpha (b/x)^{\alpha}}{a + (b/x)^\alpha}\frac{\Delta B_r}{B_r}
\end{equation}
In this numerical integration -- which, we wish to stress again, represents a realistic and important case -- we find $a \ll b$ and $\alpha \approx 0.83$, therefore
\begin{equation}
\frac{\Delta t_{thr}}{t_{thr}} \approx 0.83 \frac{\Delta B_r}{B_r}
\end{equation}
so that there is a slight damping of the fluctuations of production rate, and this is an additional factor that contributes to the increased robustness of this process with multiple site phosphorylation. 

\section{Realistic phosphorylation/dephosphorylation dynamics}
\label{add}

The attachment-detachment dynamics in phosphorylation/dephosphorylation chains is actually more complex than the modification dynamics introduced in section \ref{diffeq} and used to analyze the example of the previous section. However the dynamics can be easily modified to include a more realistic attachment-detachment process, like that described in \cite{sha}, where phosphorylation and dephosphorylation proceed as follows
\[\text{protein} + \text{ATP} \overset{\text{cyclin-CDK}}{\longrightarrow} \text{protein-P} + \text{ADP}\]
\[\text{protein-P} + \mathrm{H}_2\mathrm{O} \overset{\text{phosphatase}}{\longrightarrow} \text{protein} + \text{P}\]
where it is assumed that the reactions proceed in an aqueous environment with plenty of ATP. Each reaction is actually a Michaelis-Menten step, and if we assume the usual quasi-steady-state approximation \cite{sm}, we obtain the new differential system
\begin{eqnarray}
\nonumber
\frac{d[A_{0}]}{dt} & = & -N \frac{k_{cat,P} [B^{(P)}] [A_{0}]}{K_{m,P}+[A_{0}]} + \frac{k_{cat,D} [B^{(D)}] [A_{1}]}{K_{m,D}+[A_{1}]}\\
\nonumber
&&\ldots\\
\nonumber
\frac{d[A_{n}]}{dt} & = & -n \frac{k_{cat,D} [B^{(D)}] [A_{n}]}{K_{m,D}+[A_{n}]}  -(N-n) \frac{k_{cat,P} [B^{(P)}] [A_{n}]}{K_{m,P}+[A_{n}]} \\
\nonumber
&& +(N-n+1) \frac{k_{cat,P} [B^{(P)}] [A_{n-1}]}{K_{m,P}+[A_{n-1}]} + (n+1) \frac{k_{cat,D} [B^{(D)}] [A_{n+1}]}{K_{m,D}+[A_{n+1}]}\\
\label{gensysMM}
&&\ldots\\
\nonumber
\frac{d[A_{N}]}{dt} & = & -N \frac{k_{cat,D} [B^{(D)}] [A_{N}]}{K_{m,D}+[A_{N}]} + \frac{k_{cat,P} [B^{(P)}] [A_{N-1}]}{K_{m,P}+[A_{N-1}]} 
\end{eqnarray}
where the $k_{cat}$'s and the $K_m$'s are the Michaelis-Menten parameters, the superscripts P and D denote respectively the phosphorylation and the dephosphorylation step,  and $[B^{(P)}]$ and $[B^{(D)}]$ are the initial (total) concentrations of the phosphorylating and of the dephosphorylating enzyme (the kinase and the phosphatase in the scheme of ref. \cite{sha}).
Notice that now the B's no longer depend on the attachment-detachment dynamics, and the differential system (\ref{gensysMM}) seems to be essentially different from the original differential system (\ref{gensys}). However the actual values of the $K_m$'s are usually large with respect to the expected protein concentrations \cite{sha} (for a simple estimate of the protein concentrations inside a cell see, e.g.,  \cite{ourpap}), and therefore from  (\ref{gensysMM}) we obtain the linear system 
\begin{eqnarray}
\nonumber
\frac{d[A_{0}]}{dt} & \approx & -N k_P [A_{0}][B^{(P)}] + k_D [A_{1}]\\
\nonumber
&&\ldots\\
\nonumber
\frac{d[A_{n}]}{dt} & \approx & -n k_D  [A_{n}] -(N-n) k_P [A_{n}][B^{(P)}]  \\
\label{linsysMM}
&&+(N-n+1) k_P [A_{n-1}] [B^{(P)}]+ (n+1) k_D [A_{n+1}]\\
\nonumber
&&\ldots\\
\nonumber
\frac{d[A_{N}]}{dt} & \approx & -N k_D [A_{N}] + k_P [A_{N-1}][B^{(P)}]
\end{eqnarray}
where $k_D = k_{cat,D}[B^{(D)}]/K_{m,D}$ and $k_P = k_{cat,P}/K_{m,P}$. It is easy to see that the system matrix is just like (\ref{mat0}), with the substitution $\bar{r} \rightarrow k_P[B^{(P)}]/k_D$, and therefore the eigenvalues are those calculated in section \ref{eigen}, i.e., 
\begin{equation}
\lambda_k = - \left\{ (N-k) + (N-k) \left[\frac{k_P[B^{(P)}]}{k_D} \right]\right\}
\end{equation}
($0 \le k \le N$): thus we see that the more realistic attachment-detachment process yields basically the same dynamics in the linear approximation.

\section{Cell-cycle control revisited}
\label{cyclecontrol2}

\begin{figure}
\centering
\includegraphics[width=3.2in]{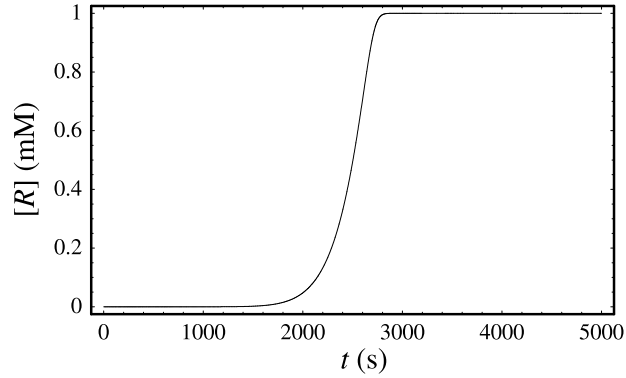}
\caption{\label{fig13} This figure shows the concentration $[R]$ of the product of the downstream Michaelis-Menten reaction of section \ref{cyclecontrol2}, when we assume that the concentration of the Cylin-CDK complex $B_P$ that phosphorylates the pRb protein grows linearly in time. The parameters are given in section \ref{cyclecontrol2} and the $B$ production rate is $1.39\cdot 10^{-11}$ M s$^{-1}$.}
\end{figure}

\begin{figure}
\centering
\includegraphics[width=3.2in]{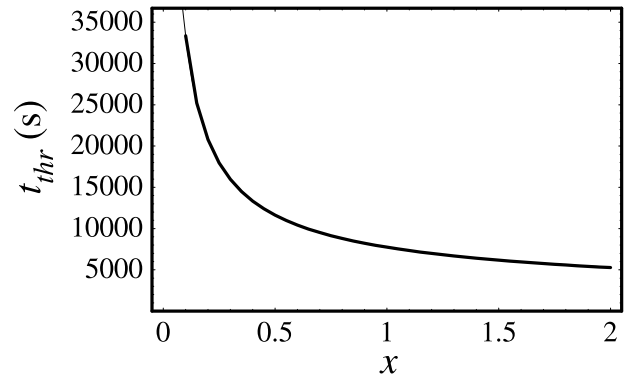}
\caption{\label{fig14} Threshold time (defined in figure \ref{fig11}) vs. $B$ production rate, from the numerical integration of section \ref{cyclecontrol2}; here the production rate is $x \cdot (1.39\cdot 10^{-11}$ M s$^{-1})$. Again, the curve is very well fit by the function  $t_{thr}(x) = a + (b/x)^\alpha$, where the exponent is $\alpha = 0.72$ (once again, fit and numerical results are so close that they are indistinguishable from each other).  }
\end{figure}

The considerations in the previous section suggest that the inclusion of the more realistic phosphorylation/dephosphorylation steps in the cascade of section \ref{cyclecontrol} should not change appreciably the numerical integration. Unfortunately, to the best of our knowledge, the actual Michaelis-Menten parameter values have never been measured and we have taken simple estimates based on related measurements \cite{pan,bart}, i.e., $k_{cat,P} \approx 0.001$ s$^{-1}$, $k_{cat,D} \approx 0.0025$ s$^{-1}$, $K_{m,P} \approx 0.92$ $\mu$M, $K_{m,D} \approx 0.94$ $\mu$M. Using these values, in addition to those already listed in section \ref{cyclecontrol}, we have integrated numerically the differential system (\ref{gensysMM}): even  though we have taken $K_m$'s that are not very large with respect to the concentrations $[A_n]$, the results are very similar to those found in section \ref{cyclecontrol}. Figure \ref{fig13} shows a single curve for the product R of the downstream Michaelis-Menten reaction, when we assume that the concentration of the Cylin-CDK complex $B_P$ that phosphorylates the pRb protein grows linearly in time (this is reasonable, see, e.g., \cite{ghina}). The $B$ production rate is assumed to be $1.39\cdot 10^{-11}$ M s$^{-1}$: this curve bears a striking similarity with those of figure \ref{fig11}. 
Finally \ref{fig14} shows the threshold time vs. the $B$ production rate: again, the curve is very well fit by the function  $t_{thr}(x) = a + (b/x)^\alpha$, where the exponent is now $\alpha = 0.72$. The power-law behavior is the same as that found in section \ref{cyclecontrol} and the exponent is also very close to that found earlier; the different value of the exponent is obviously due to the more complete dynamics. 
These results indicate that the actual attachment-detachment dynamics is not important and that the allosteric effect is independent of its details.

\section{Conclusions}
\label{concl}

In this paper we have produced a detailed study of the dynamics of multisite protein modification, and have analyzed both the equilibrium and the dynamical properties of the system. This paper follows a previous work \cite{ourpap} where we have shown that multisite protein modification may be used by cells both to set the time scale of a process -- changing it by orders of magnitude -- and to make it more robust against environmental and endogenous sources of variability. Initially we have isolated the attachment/detachment dynamics and have explicitly calculated the relaxation rates (eigenvalues) of the linearized differential system, and thus the characteristic time scales. The system matrix has an interesting, nearly ordered shape and -- to the best of our knowledge -- the eigenvalues that we find here were previously unknown. We have also used these rates to compute the synthetic noise spectra, which are relevant when the number of molecules is small and discreteness plays an important role. We have extended these results with numerical calculations, and we find that possible memory effects, that show up as power laws in noise spectra (and here we recall that the higher the spectral index, the greater the correlation between individual modification events), are suppressed when the modification chain is coupled to a threshold process, and the approach to equilibrium is fast.
We have considered a process which is very similar to the chain of reactions that leads to the crucial restriction point in the cell cycle, and we have shown that multisite phosphorylation acts in this case as a threshold stabilizing factor, that helps reduce individual differences between cells that may show up as a different cyclin synthesis rate,  and thus stabilizes the duration of the cell cycle. These numerical results have been obtained with a grossly simplified phosphorylation/dephosphorylation dynamics, and for this reason we have considered next a more realistic dynamics. We find that the conclusion obtained in the section on the linearized dynamics still hold, and moreover an explicit numerical integration of the more realistic dynamics yields essentially the same results as the simpler bimolecular attachment-detachment dynamics: this indicates that the allosteric effect is not actually dependent on the details of the modification process.

\newpage

%figures

\end{document}